\RequirePackage{ifpdf}
\ifpdf 
\documentclass[pdftex]{sigma}
\else
\documentclass{sigma}
\fi

\begin{document}

\numberwithin{equation}{section}

\def\cte{\alpha}
\def\cteb{\gamma}
\def\ctec{\gamma}
\def\otra{b}
\def\co{\Delta}
\def\pois#1#2{\left\{ {#1},{#2} \right\}}
\def\la{\lambda}

\allowdisplaybreaks

\renewcommand{\PaperNumber}{033}

\renewcommand{\thefootnote}{$\star$}

\FirstPageHeading

\ShortArticleName{Continuous and Discrete (Classical) Heisenberg
Spin Chain Revised}

\ArticleName{Continuous and Discrete (Classical)\\ Heisenberg Spin
Chain Revised\footnote{This paper is a contribution to the
Proceedings of the Workshop on Geometric Aspects of Integ\-rable
Systems
 (July 17--19, 2006, University of Coimbra, Portugal).
The full collection is available at
\href{http://www.emis.de/journals/SIGMA/Coimbra2006.html}{http://www.emis.de/journals/SIGMA/Coimbra2006.html}}}

\Author{Orlando RAGNISCO and Federico ZULLO}
\AuthorNameForHeading{O.~Ragnisco and F.~Zullo}

\Address{Dipartimento di Fisica,   Universit\`a di Roma Tre and
Istituto Nazionale di Fisica Nucleare\\ Sezione di Roma Tre,  Via
Vasca Navale 84, I-00146 Roma, Italy}

\Email{\href{mailto:ragnisco@fis.uniroma3.it}{ragnisco@fis.uniroma3.it},
\href{mailto:federico.zullo@virgilio.it}{federico.zullo@virgilio.it}}

\ArticleDates{Received December 29, 2006; Published online February 26, 2007}

\Abstract{Most of the work done in the past on the integrability
structure of the Classical Heisenberg Spin Chain (CHSC) has been
devoted  to studying the $su(2)$ case, both at the continuous and
at the discrete level. In this paper we address the problem of
constructing integrable  generalized ``Spin Chains'' models, where
the relevant f\/ield variable is represented by  a $N\times N$
matrix whose eigenvalues are the $N^{\rm th}$ roots of unity. To
the best of our knowledge, such an extension has never been
systematically pursued. In this paper, at f\/irst we  obtain  the
continuous $N\times N$ generalization of the CHSC  through the
reduction technique for Poisson--Nijenhuis manifolds, and exhibit
some explicit, and hopefully interesting, examples for $3\times 3$
and $4\times 4$ matrices;  then, we discuss the much more
dif\/f\/icult discrete case, where  a few partial new results are
derived and a  conjecture is made for the general case.}


\Classification{37K05; 37K10}

\section{Introduction}

\looseness=1 The Heisenberg ferromagnet has been one of the
f\/irst integrable models investigated and solved
in the framework of the Inverse Scattering Method, both at a
classical \cite{Thak,FT,Orph} and at a~quantum level. In
particular, at the quantum level, through the  Quantum Inverse
Scattering Method \cite{KS,Sklianin}, the original Bethe solution
\cite{Bethe} has been rediscovered and generalized, and the
solution of the isotropic one-dimensional spin chain  was just the
starting example in a list of a~number of outstanding
achievements obtained in this f\/ield in the 80's and in the 90's
of the last century. The scope of the present contribution is
twofold. On one hand, in Section~2  the geometric approach
introduced in the f\/ield of inf\/inite dimensional integrable
systems by Magri and coll. \cite{MM,PN, MMR} will be used to
extend to $N\times N$ matrices  the old results on the $SU(2)$
classical Heisenberg chain obtained about 20 years ago;
accordingly, new hierarchies of integrable evolution equations
will be derived, and a few explicit examples will be shown, which
might have some relevance for studying classical analogs of the
so-called Potts model \cite{Potts,Wu,Hart}. On the other hand, in
Section~3 the intriguing issues concerning the lattice version of
the model \cite{FT, RS} will be reexamined, a few additional
results in the $2\times 2$ case  will be given, and f\/inally  a
conjecture concerning the general $N\times N$ case will be
formulated.

\section[The classical Heisenberg chain on the line for  $N\times N$ matrices]{The classical Heisenberg chain on the line for  $\boldsymbol{N\times N}$ matrices}

Let us f\/irst recall the basic notions and results underlying the
integrability of the classical Heisenberg ferromagnet in the usual
$su(2)$ setting.  As a starting point, we will adopt the simplest
point of view, where an integrable (matrix) system in $1+1$
continuous dimensions is regarded as the compatibility condition
for two linear dif\/ferential (matrix) systems, def\/ining the so
called associated Lax pair:
\begin{gather}
\psi_x = \lambda S\psi, \label{Laxa}\\
\psi_t = V \psi. \label{Laxb}
\end{gather}
In (\ref{Laxa}), (\ref{Laxb})  $\psi$, $S$, $V$ are $n\times n$
matrix valued functions of the $(x,t)$ coordinates,  $S$ is the
f\/ield variable appearing linearly in the so-called principal
spectral problem, $V$ has an additional (rational) dependence on
a complex parameter $\lambda$, denoted as the spectral parameter.
It would be natural to look at  $\psi$ as an element of a  Lie
group, and at  $S$, $V$ as elements of the associated  Lie
algebra. More correctly, due to the extra dependence on
$\lambda$, we should speak of ``loop'' Groups and Algebras.
However, in the sequel  we will  always assume $S$ to belong to
the open set of the invertible elements of the unital associative
algebra  of the  $N\times N$ matrices with (generally) distinct
eigenvalues.

The compatibility condition of the above linear system
(\ref{Laxa}), (\ref{Laxb}) reads:
\begin{gather}
S_t= {\frac{1}{\la}}V_x + [V,S]\label{comp}
\end{gather}
which is also known as the zero-curvature condition for the
connection given by the one form:
\[
\alpha = \lambda S dx - Vdt.
\]

Actually, in the very simple example under scrutiny, the
compatibility condition (\ref{comp}) amounts to the compatibility
(i.e.\ vanishing of the Schouten bracket) of the two Poisson
tensors:
\begin{gather*}
{\mathcal{P}}V = [V,S], \qquad {\mathcal{Q}}V = V_x 
\end{gather*}
which then def\/ine the bi-Hamiltonian pair ${\mathcal{P}}$,
${\mathcal{Q}}$.

As is well known from \cite{MM,PN,MMR}, the Poisson tensors $
{\mathcal{P}}$, ${\mathcal{Q}}$ are linear operators from the
cotangent bundle $T^\ast M$ to the tangent bundle $TM$ to our
conf\/iguration manifold, which is just an af\/f\/ine hyperplane
to the linear space of matrix-valued $C^\infty$ functions of the
real variable $x$ fulf\/illing homogeneous boundary conditions.
The duality form between $TM$ and $T^\ast M$ is the usual trace
form  $\langle \phi,\alpha\rangle := \int_{-\infty}^\infty {\rm
Tr}\,( \phi \alpha) dx$. As ${\mathcal{Q}}$ is invertible, we can
introduce on $TM$ the  tensor (Recursion operator):
 \begin{gather}
 \mathcal{N}:=\mathcal{P}\mathcal{Q}^{-1} \ \Rightarrow \ \phi^\prime (:= S_{t^\prime}) =
 \left[S, \int^x \phi (:= S_t)\right]\label{Rec}
\end{gather}
that has zero Nijenhuis torsion (whence the name ``Nijhenhuis
tensor'' \cite{Nij}) and gives rise to a~hierarchy of commuting
f\/lows. Under the above conditions we will say that $M$ is
equipped with a ${\mathcal{P}}\mathcal{N}$ structure.

In order to construct a hierarchy of integrable evolution
equations, we have to choose a submanifold $M' \subset M$ such that
the Nijenhuis tensor has maximal rank on it. As shown
in~\cite{MMR} we can achieve this goal by  going on restricting
the action of ${\mathcal{N}}$ to  smaller and smaller submanifolds
as far as  the  structure is preserved. In particular, it is
always possible  to restrict~${\mathcal{N}}$ on  a~characteristic
leaf of ${\rm Im}\, {\mathcal{Q}}$, and then again on a  leaf  of
the distribution generated by ${\rm Im}\,{\mathcal{N}}^r$, where
$r$  (assumed to be f\/inite) is the Riesz index of
${\mathcal{N}}$ \cite{Krey}. As in our case ${\mathcal{Q}}$ is
invertible,  the only relevant restriction is on
\begin{gather}
{\rm Im}\, {\mathcal{N}} = {\rm Im}\,  {\mathcal{P}}\{ \phi : \phi
= [S,\alpha]~{\rm {for~some~\alpha}} \}. \label{Im}
\end{gather}
It can be readily seen that ${\rm Im}\, {\mathcal{N}} = {\rm Im}\,
{\mathcal{N}}^k$, ${\rm Ker}\, {\mathcal{N}} = {\rm
Ker}\,{\mathcal{N}}^k$, $\forall\, k >1$. Hence $r = 1$ and then
we can restrict the ${\mathcal{P}}\mathcal{N}$ structure on a
characteristic leaf of the distribution  (\ref{Im}):
\[ TM' = \{\phi: ~{\rm Tr}\, S^k \phi = 0\}\qquad (k=0, \dots, N-1)
\]
yielding the symplectic leaves:
\begin{gather}
{\rm Tr}\, S^k = d_k. \label{symp}
\end{gather}

The hierarchy obtained by acting with $\mathcal{N}$ (\ref{Rec})
on the starting symmetry $[S,C]$, $C$ being an arbitrary constant
matrix, (the so-called ``rotational f\/low'') consists however of
$nonlocal$ equations. As it was already shown  in \cite{MMR} for
$2\times 2$ matrices, the way  to get local equations is then  to
invert ${\mathcal{N}}$, which is obviously possible thanks to the
above restriction.

In the following, for the sake of simplicity, we will simply
denote by $ \mathcal{N}^{(-1)}$ the inversion of the restriction
of $ \mathcal{N}$ on (\ref{symp}).

The  most important case, and the  best studied one, is when $S
\in su(2)$, namely it is a $2\times 2$ Hermitean traceless matrix:
\[
S = {\vec{S}} \cdot \hat {\bf{\sigma}},\qquad {\vec{S}} \cdot
{\vec{S}} =1,\qquad \hat {\bf{\sigma}}
 = ( {\bf{\sigma}}_1, {\bf{\sigma}}_2, {\bf{\sigma}}_3).
 \]
Finding $\mathcal N^{-1}$ is a rather easy task. Indeed, from
\[ \phi^\prime  = \left[S, \int^x \phi\right]
\]
one gets
\[
 \phi = \frac{1}{4} [S, \phi^\prime]_x + (c_0 I + c_1S)_x,
 \]
where the second term in the above formula comes from ${\rm Ker}\,
[S, \cdot]$. Whence, imposing $\phi$ to be tangent to the
symplectic leaves (\ref{symp}),  we derive:
\[
c_0=0,\qquad c_1 = -\frac{1}{8}\int^x {\rm tr}\, [S,S_x]\phi
\]
so that:
\begin{gather} \phi : = \mathcal{N}^{(-1)}\phi^\prime =  \frac{1}{4} [S, \phi^\prime]_x  - \frac{1}{8}S\int^x {\rm tr}\, [S,S_x]\phi. \label{Ninverso}
\end{gather}
From (\ref{Ninverso}) one gets a local hierarchy \cite{MMR}, whose
f\/irst members are:
\[
 \phi_0 =0, \qquad \phi_{-1} = S_x,\qquad  \phi_{-2} = [S,S_{xx}],\quad \dots.
 \]

In particular the  HSC Hamiltonian f\/low is $\phi_{-2} =
[S,S_{xx}]$, namely:
\begin{gather}
S_t = [S,S_{xx}] = [S, \nabla {\mathcal{H}}] \quad \Rightarrow
\quad {\mathcal{H}} = {\frac{1}{2}}\int_{-\infty}^{+\infty} {\rm
Tr}\, (S_x)^2. \label{HM}
\end{gather}

The novel result contained in this section is the explicit form of
$\mathcal{N}^{(-1)}$, and of the associated local evolution
equations and Hamiltonian functionals for  $N\times N$ matrices
s.t.\ $S^N =I$, which we have denoted as $ HSC$ {\it{at the
$N^{\rm th}$ roots of unity}}. The derivation is quite simple.
Indeed, from the very def\/inition of the Nijhenhuis tensor
(\ref{Rec}) one easily gets:
\[
 \phi = \frac{1}{2N} \frac{\partial}{\partial x}
 \left(\sum_{l=1}^{N-1}l\big[S^l\phi^\prime S^{N-l-1} - S^{N-l-1} \phi^\prime S^l\big]+ \sum_{l=0}^{N-1}c_l S^l\right).
\]
The conditions:
\[
{\rm Tr}\, S^k\phi= 0
\]
 that ensure $\phi$ to be tangent to (\ref{symp}),
 allow to determine  $c_k$ through quadratures, up to constant integration factors:
\[
(c_k)_x = -\frac{2}{N}\,{\rm Tr}\sum_{l=1}^{N-1}l
S^{n-l-1}\phi^\prime S^l (S^{N-k})_x.
\]

For a given $N$, the hierarchy of evolution equations splits into
$N-1$ sub-hierarchies, with starting symmetries $(S^k)_x$
($k=1,\dots,N-1$). In the following, we provide the explicit form
of the f\/irst members of the hierarchy  for $N=3$ and $N=4$.

\begin{example} $N=3$.
 There are two families of local f\/lows, with starting points:
\[
\phi_1^{(1)} = S_x,\qquad \phi_1^{(2)} = (S_x)^2
\]
 whence the f\/irst few members of simpler family read:
\[
\phi_2^{(1)} = [S^2,S_x]_x, \qquad \phi_3^{(1)}=
[S^2,[S^2,S_x]_x]_x, \quad \dots.
\]
 The Hamiltonian ${\mathcal{H}}_2^{(1)}$ reads (up to constant factors):
\[
{\mathcal{H}}_2^{(1)} = \int_{-\infty}^{+\infty} {\rm Tr}\,
S(S_x)^2.
\]
\end{example}

\begin{example} $N=4$.
There are three families of local f\/lows, with starting points:
\[
\phi_1^{(1)} = S_x,\qquad \phi_1^{(2)} = (S_x)^2,\qquad
\phi_1^{(3)}  =(S_x)^3
\]
whence  the next f\/low in the simplest family reads:
\[
\phi_2^{(1)} = 3[S^3,S_x]_x + [S,SS_xS]_x,
\]
while the associated Hamiltonian functional
${\mathcal{H}}_2^{(1)}$  is given by:
\[
{\mathcal{H}}_2^{(1)} = \int_{-\infty}^{+\infty} {\rm Tr}\,
\left((S_x)^2-\tfrac{1}{8}(SS_x)^2\right).
\]
\end{example}

\begin{remark}
The f\/low (\ref{HM}) is obviously consistent with the restriction
$S^N = I$ $\forall\, N$. However, it belongs to the integrable
hierarchies only for $N=2$. In fact, the Lie derivative of the
Nijenhuis tensor (\ref{Rec}) in the direction of $\psi:=
[S,S_{xx}]$ on an arbitrary vector $\phi$ can be calculated from
the formula:
\[
{\mathcal{L}}_\psi ({\mathcal{N}}\phi) = ({\mathcal{L}}_\psi
({\mathcal{N}})(\phi) + {\mathcal{N}}{\mathcal{L}}_\psi \phi
\]
whence, with the proper replacements:
\begin{gather}
({\mathcal{L}}_\psi ({\mathcal{N}})(\phi) =
3[S,[\phi,S_x]].\label{Lie}
\end{gather}
From (\ref{Lie}) we readily see  that the Lie derivative of the
Nijenhuis tensor in the direction (\ref{HM}) vanishes only for the
subhierarchies starting with  $\phi_1^{(1)} = S_x$. Hence, for $N$
larger than~2, the f\/low (\ref{HM}), though being Hamiltonian and
possessing inf\/initely many conserved quantities (and commuting
symmetries) is certainly not completely integrable.
\end{remark}

 \section{The discrete case: some results and a conjecture}
In contrast with the continuous case, we cannot give any new
result for arbitrary $N\times N$ matrices. However, before
focussing on the well known  $2\times 2$ case, we will write down
the compatibility scheme and  recall the abstract
${\mathcal{P}}{\mathcal{N}}$ structure in the
dif\/ferential-dif\/ference setting, as it has been  elucidated
in the past by various authors \cite{RS,MT1,MT2}. The two relevant
linear problems read:
\begin{gather}
 \psi_{m+1} =(I+ \la S_m)\psi_m,\nonumber\\
\psi_{m,t} = V_m(\la)\psi_m.\label{Dlax}
\end{gather}
In (\ref{Dlax}) $\psi$, $S$, $V$ are $N\times N$ matrices,
depending on a discrete variable $m\in {\mathbb{Z}}$,  on a
continuous variable $t\in {\mathbb{R}}$ and on a complex parameter
$\lambda$. It has been shown \cite{RS,MT1,MT2} that the
compatibility of  the above linear problems yields the
${\mathcal{P}}{\mathcal{N}}$ structure:
 \begin{gather}
 \phi := {\mathcal{P}}\alpha = [S,\alpha],\nonumber \\
 \phi^\prime: = N\phi  = \frac{1}{2}[S_, ({\bf{R}}\phi)]
+ \frac{1}{2}\{S\phi\}, \label{DPN}
\end{gather}
where:
\[
{\bf{R}} = \frac {{\mathcal{ E} }+1}{{\mathcal{ E} }-1},\qquad
({\mathcal{ E}}\phi)_m := \phi_{m+1}.
\]
We note that ${\bf{R}}$ is the so-called  ILW operator, satisfying
the modif\/ied Yang--Baxter equation
 in the algebra of shift operators \cite{MT1}.

\begin{remark}
Notice that now ${\rm Im}\, {\mathcal{N}}$ is larger than ${\rm
Im}\,  {\mathcal{ P}}$; however ${\rm Im}\, {\mathcal{ P}}$ is
still an {\it{invariant submanifold}}  for ${\rm Im}\,
{\mathcal{N}} $, namely  ${\rm Tr}\, S^l\phi = 0$ implies ${\rm
Tr}\, S^l{\mathcal{ N}}\phi = 0$ $\forall\, l$. Indeed, from
(\ref{DPN}) it follows the recurrence relation ${\rm Tr}\, S^l
{\mathcal{ N}}\phi = {\rm Tr}\, S^{l+1}\phi $, entailing
 ${\rm Tr}\, [{\mathcal{ N}}^N-{\mathcal{I}}]\phi=0= {\rm Tr}\, S^l[{\mathcal{ N}}^N-{\mathcal{I}}]\phi$, $\forall\, l$.
 \end{remark}

As we mentioned before,  to the best of our knowledge explicit
evolution equations can be found in the literature only for
$2\times 2$ matrices. They are for instance reported in the
fundamental reference \cite{FT}, where the corresponding ``Lax
pair" (\ref{Dlax}) is also shown.
 As it was f\/irst pointed out in \cite{RS}, both the Discrete Heisenberg (DH) ferromagnet and the Discrete Translational (DT) f\/low  belong to the kernel of ${\mathcal{ N}}^2-{\mathcal{I}}$, which is in fact the linear span of them. They read respectively:
\begin{gather*}
 S_{m,t} = [S_m, \beta_m S_{m+1} + \beta_{m-1} S_{m-1}],\nonumber \\
 S_{m,\tau} = \beta_m (S_m + S_{m+1}) - \beta_{m-1} (S_m + S_{m-1}),
 \end{gather*}
 where
$\beta_m := \frac {1}{{\rm Tr}\, (I+S_mS_{m+1})}$. Each   of the
above two equations can be derived   from the other by applying
the recursion operator ${\mathcal{N}}$.

The above results, and Remark 2 strongly suggest the following

\medskip

\noindent {\bf Conjecture.} {\it For $N\times N $ matrices obeying
the cyclic constraint $S^N=I$, the local evolution equations span
${\rm Ker}\,  ({\mathcal{ N}}^N-{\mathcal{I}})$; hence,  starting
from any of them, the others can be obtained by applying
subsequently ${\mathcal{ N}}$. In other words, we can always
f\/ind in ${\rm Ker}\, ( {\mathcal{ N}}^N-{\mathcal{I}})$ a cyclic
basis~$\phi^{(r)}$ such that: $\phi^{(r+1)} =
{\mathcal{N}}\phi^{(r)}$, $\phi^{(r+N)}=\phi^{(r)}$.}

\subsection*{Acknowledgements}

O.~Ragnisco would like to thank the organizers of the Geomis
workshop, and in particular Joana Nunes da Costa, for the
admirable work they have done in preparing and directing the
meeting, and for their  kind and warm hospitality in Coimbra.
Also, O.R. acknowledges  illuminating discussions with his
long-time colleague and friend Franco Magri.

\pdfbookmark[1]{References}{ref}
\LastPageEnding

\end{document}